\documentclass[twocolumn,showpacs,preprintnumbers,amsmath,amssymb,prl]{revtex4}

\usepackage{graphicx}
\usepackage{dcolumn}
\usepackage{epsf}

\begin{document}

[ cond-mat/0201251, \qquad Phys. Rev. Lett. (accepted)

\title{Flux flow of Abrikosov-Josephson vortices along grain boundaries in 
high-temperature superconductors}
\author{A. Gurevich$^{1}$, M.S. Rzchowski$^{1,2}$, G. Daniels$^{1}$, S. Patnaik$^{1}$, 
B.M. Hinaus$^{3}$, F Carillo$^{4}$, F. Tafuri$^{4}$, and D.C. Larbalestier$^{1}$.} 
\address{$^{1}$Applied Superconductivity Center, 
University of Wisconsin, Madison, Wisconsin} 
\address{$^{2}$ Department of Physics, 
University of Wisconsin, Madison, Wisconsin} 
\address{$^{3}$Department of Physics, University of Wisconsin, Stevens Point, Wisconsin}
\address{$^{4}$Universita di Napoli Federico II, Dipartamento di Scienze Fisiche, Italy}

\date{\today}

\begin{abstract}
We show that low-angle grain boundaries (GB) in high-temperature superconductors exhibit 
intermediate Abrikosov vortices with Josephson cores, whose length $l$ along GB is smaller that 
the London penetration depth, but larger than the coherence length. 
We found an exact solution for a periodic vortex structure moving along GB in a magnetic 
field $H$ and calculated the flux flow resistivity $R_F(H)$, and the nonlinear voltage-current 
characteristics.  The predicted $R_F(H)$ dependence describes well our experimental data 
on $7^{\circ}$ unirradiated and irradiated $YBa_2Cu_3O_7$ bicrystals,  from which the 
core size $l(T)$, and the intrinsic depairing density $J_b(T)$  
on nanoscales of few GB dislocations were measured for the first time. 
The observed temperature dependence of  $J_b(T)=J_{b0}(1-T/T_c)^2$ indicates a significant order parameter 
suppression in current channels between GB dislocation cores.   
    
\end{abstract}
]
\maketitle

Mechanisms of current transport through grain boundaries (GB) in high-temperature 
superconductors (HTS) have attracted much attention, because a GB is a convenient  
tool to probe the pairing symmetry of HTS by varying 
the misorientation angle $\vartheta$ between the neighboring crystallites \cite{sym,gb}. 
As $\vartheta$ increases, the spacing between the GB dislocations  
decreases, becoming comparable to the coherence length $\xi(T)$ at the angle 
$\vartheta_0\simeq 4-6^{\circ}$.  The exponential decrease of the GB critical current 
density $J_b=J_0\exp(-\vartheta/\vartheta_0)$ \cite{gb}, 
makes GBs one of the principal factors limiting critical currents of HTS \cite{dcl}. 
Atomic structure of GBs revealed by high-resolution electron microscopy 
have been used to determine local underdoped states of GB, defect-induced suppression of 
superconducting properties at the nanoscale and controlled increase of $J_b$ by overdoping of 
GB\cite{gb,dop}. Much progress has been made in understanding the microscopic factors 
controlling $J_b(\vartheta)$ at zero magnetic field, but the behavior of vortices on 
GBs is known to much lesser extent.  

The extreme sensitivity of $J_b(\vartheta)$ to the 
misorientation angle makes GB a unique tool to trace the fundamental transition 
between Abrikosov (A) and Josephson (J) vortices\cite{ag} in a magnetic field $H$ 
above the lower critical field $H_{c1}$. For $\vartheta\ll \vartheta_0$, vortices on a GB 
are A vortices with normal cores pinned by GB dislocations \cite{diaz}.  For $\vartheta>\vartheta_0$, 
the maximum vortex current density circulating across the GB is limited to its {\it intrinsic} 
$J_b(\vartheta)$, much smaller then the bulk depairing current density $J_d$.  Because vortex 
currents must cross the GB which can only sustain $J_b\ll J_d$, the normal core of an A vortex 
turns into a J core, whose length $l\simeq \xi J_d/J_b$ along the GB
is greater then $\xi$, but smaller then the London penetration depth $\lambda$, if 
$J_b>J_d/\kappa$, where $\kappa=\lambda/\xi\simeq 10^2$ \cite{ag}. 
As $\vartheta$ increases, the core length $l(\vartheta)\simeq\xi J_d/J_b(\vartheta)$ increases, so  
the GB vortices evolve from A vortices for $\vartheta\ll\vartheta_0$ to mixed Abrikosov vortices 
with Josephson cores (AJ vortices) at $J_d/\kappa<J_b(\vartheta)<J_d$. The AJ vortices turn into J vortices 
at higher angles, $\vartheta >\vartheta_J\simeq \vartheta_0\ln(\kappa J_0/J_d)$, for which  
$l(\vartheta)$ exceeds $\lambda$. For $\vartheta_0=5^{\circ}$, $\kappa=100$, and $J_0=J_d$, the AJ vortices 
determine the in-field behavior of GBs in the crucial region $0<\vartheta<\vartheta_J\simeq 23^{\circ}$ 
of the exponential drop of $J_b(\vartheta)$ (in a film of thickness $d\ll \lambda$, the AJ region 
$\vartheta<\vartheta_J\simeq \vartheta_0\ln(\lambda^2/d\xi)$ broadens even further).

The AJ structures have two length scales: the core size $l>\xi$ and the intervortex spacing 
$a=(\phi_0/B)^{1/2}$.  The larger core of AJ vortices leads to their weaker 
pinning along a GB, which thus becomes a channel for motion of AJ vortices between pinned 
A vortices in the grains \cite{ag,gc} (Fig. 1). The percolative motion of AJ vortices along GBs gives rise to a 
linear region in the $V-I$ characteristic of HTS polycrystals\cite{diaz,ornl,cad,natlab,evetts}. 
However no present experimental techniques can probe the cores of GB vortices, 
because the lack of the normal core makes AJ vortices "invisible" under STM, while neither the Lorentz 
microscopy nor magneto-optics have sufficient spatial resolution to distinguish A and AJ vortices.
In this Letter we report a combined theoretical and experimental analysis which 
enabled us to prove the existence of AJ vortices in  $7^{\circ}$  $YBa_2Cu_3O_7$ bicrystals 
and extract the core size $l(T)$, and the intrinsic depairing current density 
$J_b$ at the GB from transport measurements.  The value $J_b$ of a GB turns out to be    
much higher than its global critical current density $J_{gb}$, which is limited by self-field and pinning 
effects \cite{gc,natlab,evetts,gray,m}. The field region in which only a single AJ vortex chain 
moves along GB while the bulk A vortices remain pinned, can be considerably expanded by 
irradiation, as shown below.

\begin{figure}			
\epsfxsize= 0.75\hsize  
\centerline{
\vbox{
\epsffile{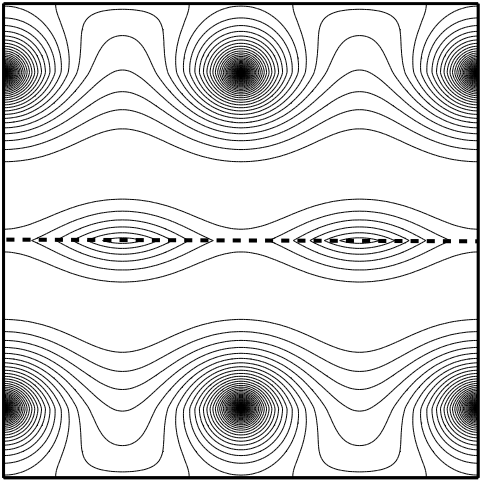} 
}}
\vskip \baselineskip
\caption{Current streamlines around AJ vortices on a GB (dashed line) and 
the bulk A vortices in the grains, calculated from Eq. (5) for $l=0.2a$.
}
\label{fig.1}
\end{figure}

For $H\gg H_{c1}$, both $l=\xi J_d/J_b$, and $a=(\phi_0/B)^{1/2}$ are smaller than $\lambda$, thus
the AJ vortices on low-angle GBs are described by a nonlocal equation for the phase 
difference $\theta(x,t)$ in the overdamped limit \cite{ag,gc}:    
	\begin{eqnarray}
	\tau\dot\theta&=&\frac{l}{\pi}\int_{-\infty}^{\infty}
	\frac{\theta^{\prime}(u)du}{u-x}-\sin\theta +\beta, 
	\label{nj} \\
	l&=&c\phi_0/16\pi^2\lambda^2J_b , 
	\qquad\qquad\tau = \phi_0/2\pi cRJ_b, 
	\label{const}
	\end{eqnarray}
where the overdot and the prime denote differentiation with respect to the time $t$ and the coordinate 
$x$ along GB, $R$ is the quasiparticle resistance of GB per unit area, $\phi_0$ is the flux quantum, 
$c$ is the speed of light, $\beta=J/J_b$, and $J(x)$ is the current density 
through GB induced by A vortices.  
Here $\beta = \beta_0 + \delta\beta (x)$ is a sum of the constant transport current 
$\beta_0$ and an oscillating component $\delta\beta(x)$ due to the discreteness of the 
A vortex lattice. The term $\delta\beta(x)$ gives rise to a critical current 
$J_{gb}$ of the GB due to pinning of AJ vortices by A vortices in the grains \cite{gc,temp}.  
Eqs. (\ref{nj}) and (\ref{const}) are independent of the pairing symmetry (which only affects $J_b$) 
and are valid for both bulk samples and thin films in a perpendicular field. 

We consider a rapidly moving AJ structure in the flux flow state, $\beta\gg\beta_c$, 
for which the pinning term $\delta\beta(x)\ll 1$ can be neglected\cite{temp}.  
In this case Eq. (\ref{nj}) has the following {\it exact} solution that describes a stable periodic 
vortex structure moving with a constant velocity $v$:
	\begin{eqnarray}
	 \theta=\pi+\gamma+2\tan^{-1}[M\tan k(x-vt)/2], \label{theta} \\
	s^2=[\sqrt{(1-\beta_0^2+h)^2+4\beta_0^2h}-1-h+\beta_0^2] /2h
	\label{sv}	
	\end{eqnarray}
Here $s=v/v_0$, $v_0=l/\tau$, $\tan\gamma = -s$, $h = (kl)^2$, $M=[1+1/h(1+s^2)]^{1/2}+[h(1+s^2)]^{-1/2}$,  
$k= 2\pi/a$, and $a$ is the period of  
the AJ structure.  For $a/l\to\infty$, Eq. (\ref{theta}) describes a moving chain of  
single AJ vortices \cite{ag}. Generally, $a(H)$ is different 
from the period of the A lattice, but for $H\gg H_{c1}$, 
the spacing $a= (\phi_0/H)^{1/2}$ is fixed by the flux 
quantization condition. Eq. (\ref{theta}) corresponds to the following field 
distribution $H(x,y)$ produced by AJ vortices in the region $|y|<\lambda$: 
	\begin{equation}
	H=\frac{\phi_0}{2\pi\lambda^2}\mbox{Re}\ln\sin [x-vt+i(|y|+y_0)]\frac{k}{2},
	\label{he} 
	\end{equation}
where $\sinh ky_0=\sqrt{h(1+s^2)}$. Eq. (\ref{he}) satisfies the Maxwell equation 
$\nabla^2H=0$ with the boundary condition $H'=(4\pi /c)[J_b\sin\theta-\hbar v\theta'/2eR-J]$ on GB, 
where $\theta (x-vt)$ is given by Eq. (\ref{theta}). 
Fig. 1 shows the current streamlines calculated from Eq. (\ref{he}). For $y>0$, these streamlines  
coincide with those of a chain of moving fictitious A vortices displaced by $y=-y_0$ away from GB. 

\begin{figure}			
\epsfxsize= 0.8\hsize  
\centerline{
\vbox{
\epsffile{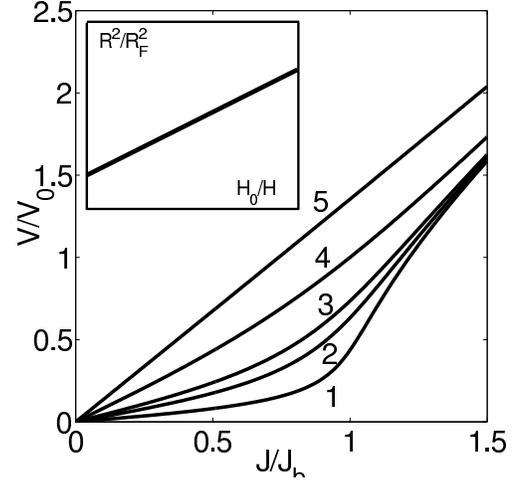} 
}}
\vskip \baselineskip
\caption{The $V-J$ curves calculated from Eq. (6) for different magnetic fields $h=H/H_0$: 
$0.01$(1), $0.05$(2), $0.1$(3), $0.5$(4), $10$(5). 
Inset shows the field dependence of $R_F(H)$.
}
\label{fig.2}
\end{figure}

The mean voltage V on a GB is determined by the Faraday law, $V=\phi_0v/ca$, which yields  
	\begin{equation}
	V= V_0\Bigl[\sqrt{(1-\beta_0^2+h)^2+4\beta_0^2h}-1-h+\beta_0^2\Bigr]^{1/2},
	\label{vj}
	\end{equation}
where $V_0=RJ_b/\sqrt{2}$. The $V-J$ curve shown in Fig. 2 is similar to that obtained 
by molecular dynamic simulations of incommensurate vortex channels \cite{kes}. In our case 
the nonlinearity of $V(J)$ is due to the AJ core expansion as $J$ increases\cite{ag}.
For $J\ll J_b$, the V-J curve is linear, $V=R_FJ$, where $R_F=R\sqrt{h/(1+h)}$ is 
the flux flow resistivity due to the viscous motion of AJ vortices. If $H\gg H_{c1}$, 
then $h= (2\pi l/a)^2=H/H_0$, and  
	\begin{equation}
	R_F=\frac{R\sqrt{H}}{\sqrt{H+H_0}},\quad\quad H_0=\frac{\phi_0}{(2\pi l)^2}
	\label{r}
	\end{equation}    
At $H\ll H_0$, Eq. (\ref{r}) describes $R_F(H)$ for AJ vortices, whose cores do not 
overlap. In this case $R_F(H)$ is reminiscent of the 1D Bardeen-Stephen formula, 
$R_{BS}\simeq R\sqrt{H/H_{c2}}$, except that in 
Eq. (\ref{r}) the core structure is taken into account exactly. For $H>H_0\simeq (J_b/J_d)^2H_{c2}\ll H_{c2}$, 
the AJ cores overlap, and Eq. (\ref{r}) describes a crossover to a field-independent quasiparticle 
resistance of GB. This regime has no analogs for A vortices, whose normal cores overlap only at $H_{c2}$. 
The simplicity of Eq. (\ref{r}) enabled us to extract intrinsic GB properties from the 
measurements of $R_F(H)$. 

\begin{figure}			
\epsfxsize= 0.8\hsize  
\centerline{
\vbox{
\epsffile{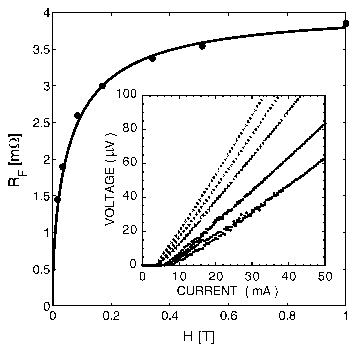} 
}}
\vskip \baselineskip
\caption{$R_F(H)$ data points extracted from the slopes of the 
$V(J)$ curves at 77K and different  H for unirradiated $7^{\circ}$ bicrystal.
The solid curves are described by Eq. (7) with $R= 4.05 m\Omega$ and $H_0=0.14 T$. 
Inset shows $V-J$ curves for $0.17$kOe, $0.51$kOe, $0.85$kOe, $1.7$kOe, and $3.4$kOe 
(from bottom to top, respectively).
}

\label{gunn}
\end{figure}

We observed the AJ vortex behavior on $7^{\circ}$ $YBa_2Cu_3O_7$,  bicrystals with a 
sharp resistive transition $\Delta T<0.4K$ at $T_c=91$K. Thin films of thickness 250 nm 
were grown on [001]-oriented $SrTiO_3$ bicrystals by pulsed laser deposition at 210 mTorr 
oxygen pressure and $810^{\circ}$C, and then annealed in oxygen at 830 Torr and 
$520^{\circ}$C for 30 min. One $7^{\circ}$ bicrystal was irradiated with 1GeV Pb ions at a fluence 
corresponding to 1T. Bridges 25 $\mu m$ wide were patterned by  Ar ion beam etching on a cooled 
sample mount to produce a four-point measurement geometry, as described in Ref. \cite{cad}. The 
voltage probes were 100$\mu$m apart, on either sides of the GB.
$V-I$ curves were measured in a gas-flow cryostat in fields $0<H<10$ T. 
Both samples are in the transition region of $\vartheta$
between strongly coupled bicrystals that behave in a manner similar to the 
single-crystal grains, and weakly coupled bicrystals that demonstrate the Josephson effect\cite{m}. 
The intragrain $J_c$ values at 77 K were $0.1 MA/cm^2$ and $0.27 MA/cm^2$ for the 
unirradiated and irradiated samples, respectively. 

Fig. 3 shows V-I characteristics for the unirradiated sample at 77K and 
different magnetic fields. The V-I curves exhibit 
linear flux flow portions in a wide range of $I$ above the depinning critical current 
$I_{gb}(H)$ which decreases with H \cite{m}. In this work we focus 
on the flux flow region $I\gg I_{gb}$, where the dynamic resistance 
$R_F(H)=dV/dI$ increases as $\sqrt{H}$ at low H, but levels off at higher 
H. We found that Eq. (\ref{r}) describes the observed $R_F(H)$ very well, thus    
the vortex cores on this GB overlap at $H\sim H_0$, well below $H_{c2}$. 
Because the GB can sustain a finite supercurrent $I_{gb}$ even for $H>H_0$,
the GB vortices lack normal cores. 
The good fit in Fig. 3 enabled us to extract $R=4.05 m\Omega$ and $H_0=0.14 T\ll H_{c2}$ 
by plotting $1/R_F^2(H)$ versus $1/H$, as shown in Fig. 2.  
Using Eq. (\ref{r}), we obtain that $l =(\phi_0/H_0)^{1/2}/2\pi = 190 \AA$ at 77K, 
thus vortices on this $7^{\circ}$ GB are indeed AJ vortices with phase cores 
much greater than $\xi(T)=\xi_0/\sqrt{1-T/T_c}\simeq 40\AA$ 
but smaller than $\lambda= \lambda_0/\sqrt{1-T/T_c}\simeq 4000\AA$  
for $\xi_0\simeq 15\AA$, $\lambda_0\simeq 1500\AA$, and $T_c=91$K.

\begin{figure}			
\epsfxsize= 0.85\hsize  
\centerline{
\vbox{
\epsffile{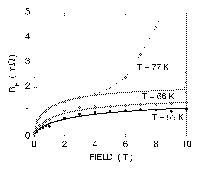} 
}}
\vskip \baselineskip
\caption{$R_F(H)$ extracted from the slopes of  
$V(J)$ for different T and H for the irradiated $7^{\circ}$ bicrystal. 
The $V-J$ curves have the extended linear regions similar to those for the unirradiated sample 
shown in Fig. 3. The solid curves are described by Eq. (7) with two fit 
parameters $R$ and $H_0$. 
}

\label{gunn}
\end{figure}

$R_F(H)$ data for the irradiated bicrystal at different $T$ are shown in Fig. 4. The good agreement 
between Eq. (\ref{r}) and the observed $R_F(H)$ enabled us to extract the temperature dependences 
of $H_0$ and $R$ shown in Fig. 5. While $R(T)$ varies only weakly, the field $H_0(T)$ exhibits a 
parabolic dependence, $H_0(T)= H_0(0)(1-T/T_c)^2$ with $H_0(0)\approx 42$T. As follows from 
Eq. (\ref{r}), the fact that $H_0(T)\propto (T_c-T)^2$ implies $l(T)\propto (T_c-T)^{-1}$, which, 
in turn, indicates the SNS behavior of $J_b(T)\propto (T_c-T)^2$, if $\lambda(T)\propto (T_c-T)^{-1/2}$ 
in Eq. (\ref{const}). From the data in Fig. 5, we can also obtain the absolute value of the intrinsic  
depairing current density $J_b$ averaged over the Josephson core length $l$. 
To do so, we write Eq. (\ref{r}) in the form $J_b=(3J_d/4)\sqrt{6\pi H_0/H_{c2}}$, which express 
$J_b$ in terms of the measured parameters $H_0$, $H_{c2}=\phi_0/2\pi\xi^2$, 
and $J_d=c\phi_0/12\sqrt{3}\pi^2\lambda^2\xi$. For $H_{c2}(T)=H_{c2}'(T_c-T)$, 
this yields $J_b\simeq (3J_d/4)[6\pi H_0(0)/T_cH_{c2}']^{1/2}(1-T/T_c)^{1/2}$, 
whence $J_b(85K)\simeq 0.3J_d(85K)$ for $H_0(0)=42T$, and $H_{c2}'=2T/K$. Likewise we get 
$J_b(77K)\simeq 0.23J_d(77K)$ for the unirradiated sample, $(H_0(77K)=0.14T)$.
Therefore, our data indicate a significant suppression of the order parameter, 
even on the low-angle $7^{\circ}$ GB, in agreement with the model of Ref. \cite{mod}.  

For the observed $H_0(T)$, the AJ core length 
$l(T)=(\phi_0/H_0(T))^{1/2}/2\pi\simeq 11(1-T/T_c)^{-1}[\AA]$, 
exceeds the bulk coherence length $\xi(T)=\xi_0/\sqrt{1-T/T_c}$ at $T_c-T\ll T_c$, but  remains smaller than 
$\lambda(T)$, except very close to $T_c$.  For instance, we obtain $l(80K)\simeq 91\AA$, 
while $\xi(80K)\simeq 43\AA$, and $\lambda(80K) = 4300\AA$. 
The length $l(T)$ also exceeds the GB dislocation spacing
$\approx 32\AA$, thus the moving AJ cores probe 
GB properties averaged over few current channels between 
dislocations.  The core length $l(\vartheta)\simeq\xi J_d/J_b(\vartheta)$ increases 
as $\vartheta$ increases, becoming larger than $\lambda$, 
if $\vartheta >\vartheta_0\ln\kappa$, in which case AJ vortices turn 
into J vortices. Since the ratio $J_b/J_d$ decreases as $T$ increases, 
higher-angle GBs can exhibit AJ vortices at low T and J vortices at $T\approx T_c$, 
while lower angle GBs have A vortices at low T and AJ vortices at $T\approx T_c$. 

\begin{figure}			
\epsfxsize= 0.8\hsize  
\centerline{
\vbox{
\epsffile{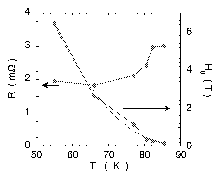} 
}}
\vskip\baselineskip
\caption{$R(T)$ and $H_0(T)$ extracted from the fit 
procedure represented in Fig. 4. The dashed line shows  
$H_0= 42(1-T/T_c)^2 [Tesla]$.
}
\label{gunn}
\end{figure}

The depinning current density $J_{gb}$ seen in Fig. 3 is due to 
interactions of AJ vortices with microstructural inhomogeneities along GB and 
pinned A vortices in the grains\cite{ag}. Another essential mechanism   
occurs if the periods of AJ and A vortices do not coincide,  
resulting in misfit dislocations in the AJ chain\cite{kes}. These vortex dislocations 
can strongly limit $J_{gb}$ because their intrinsic pinning  
by the A vortex lattice is exponentially weak\cite{kes}. For $H\gg H_{c1}$, when the 
AJ and A periods are close, the pinning of a few misfit dislocations 
in the AJ chain may be thus dominated by macroscopic $T_c$ and $J_b(x)$ variations along GB 
caused by facet structures, strain fields, local nonstoichiometry, etc.  
\cite{cai}.  In this Letter we focus 
on the flux flow state $J\gg J_{gb}$, where pinning weakly affects $R_F$. In this case 
measurements of $R_F(H)$ reveal the essential physics of GB vortices whose moving AJ cores probe  
intrinsic properties of GBs at the nanoscale, regardless of particular pinning mechanisms which 
are essential at $J\approx J_{gb}$.   
Because the observed $V-J$ curves are nearly linear above $J_{gb}(77K, 1T)\sim  
10^4-10^5 A/cm^2\ll J_b\simeq (0.2-0.3)J_d\sim 1-10 MA/cm^2$, 
the pinning region $J\sim J_{gb}$ is much smaller than the scale of Fig. 2. Thus, the 
nonlinearitry of $V(J)$ due to the AJ core expansion does not affect   
$R_F$ measured at $J<3J_{gb}$ so Eq. (\ref{r}) can be used to fit the data. A similar approach 
was used to measure the flux flow resistivity of 
pinned A vortices driven by strong current pulses well above $J_c$\cite{kunchur}.    

The case when only a single AJ vortex row moves along the GB, while the 
intragrain A vortices remain pinned corresponds to low fields, $H<H_1$. 
For $H>H_1$, the AJ vortices drag neighboring A vortices in the flux flow 
channel along GB \cite{gc}. The field $H_1$ can be estimated from the condition 
that the pinning force $f=\phi_0\Delta H/2a(H)$ of AJ vortices due to their magnetic 
interaction with A vortices equals the bulk pinning force $\phi_0J_c/c$, 
where $\Delta H = \phi_0e^{-2\pi u/a}/\pi\lambda^2$ is the amplitude of the oscillating 
part of the local field $H(x)\approx B+\Delta H\cos (2\pi x/a)$ produced by A vortices along GB, and 
$u$ is the spacing of the first A vortex row from GB\cite{temp}. Therefore,
	\begin{equation}
	H_1\simeq 4\phi_0J_c^2(H_1)/c^2\Delta H^2
	\label{h1}
	\end{equation}
Transitions from a single to a multiple row vortex motion \cite{cad,evetts} 
result in a sharp upturn of the $R_F(B,77K)$ curve at $H_1\simeq 4T$ in Fig. 4.
Here $H_1$ for our unirradiated sample ($J_c=0.1 MA/cm^2$) is 7.3  times smaller 
than $H_1$ for the irradiated one ($J_c=0.27 MA/cm^2$).
For $\lambda(77K)\simeq 4000\AA$, and $H_1=4T$, Eq. (\ref{h1}) yields  
$u=\ln(c^2H_1\phi_0/4\pi^2\lambda^4J_c^2)/4\pi=0.92a$. 

In conclusion, vortices on 
low-angle grain boundaries in HTS are mixed Abrikosov-Josephson vortices.  
Exact solutions for a moving AJ vortex structure, the nonlinear $V-J$ characteristic 
and the flux flow resistivity  $R_F(B)$ were obtained. From the measurements of $R_F(B)$ 
on a $7^{\circ}$  YBCO bicrystal, we extracted the length of the AJ core and the 
intrinsic depairing current density $J_b$ on GB. The analysis proposed in this work can 
be used for systematic studies of the effect of overdoping \cite{dop} on the local 
suppression of the order parameter and current transport through nanoscale channels 
between GB dislocations.  
  
This work was supported by the NSF  MRSEC (DMR 9214707),
AFOSR MURI (F49620-01-1-0464), and by Italian INFM-PRA JT3D.


\begin{references}

\bibitem{sym}C.C. Tsuei and J.R. Kirtley, Rev. Mod. Phys. {\bf 72}, 969 (2000).
\bibitem{gb}H. Hilgenkamp and J. Mannhart, Rev. Mod. Phys. {\bf 74},  (2002).
\bibitem{dcl}D.C. Larbalestier, A. Gurevich, D.M. Feldmann, and A.A. Polyanskii, 
Nature {\bf 413}, 368 (2001). 
\bibitem{dop}S.E. Russek {\it et al.}, Appl. Phys. Lett. {\bf 57}, 1155 (1990); 
A. Schmehl {\it et al.}, Europhys. Lett. {\bf 47}, 110 (1999); 
G. Hammerl {\it et al.}, Nature {\bf 407}, 162 (2000); K. Guth {\it et al.}, \prb{\bf 64}, R140508 
(2001).
\bibitem{ag}A. Gurevich, \prb{\bf 46}, R3187 (1992); {\bf 48}, 12857 (1993); 
Physica C{\bf 243}, 191 (1995). 
\bibitem{diaz}A. Diaz {\it et al.}, \prl {\bf 80}, 3855 (1998); \prb{\bf 58}, R2960 (1998).
\bibitem{gc}A. Gurevich and L.D. Cooley, \prb {\bf 50}, 13363 (1994). 
\bibitem{ornl}D.T. Verebelyi {\it et al.}, Appl. Phys. Lett. {\bf 76}, 1755 (2000); {\bf 78}, 2031 (2001). 
\bibitem{cad}G.A. Daniels, A. Gurevich, and D.C. Larbalestier, Appl. Phys. Lett. 
{\bf 77}, 3251 (2000). 
\bibitem{natlab}D. Kim {\it et al.}, \prb{\bf 62}, 12505 (2000); J. Albrecht {\it et al.}, \prb {\bf 61}, 12433 (2000).
\bibitem{evetts}M.J. Hogg {\it et al.}, Appl. Phys. Lett. {\bf 78}, 1433 (2001).
\bibitem{gray}K.E. Gray {\it et al.}, \prb {\bf 58}, 9543 (1998). 
\bibitem{m}R.D. Redwing {\it et al.}, Appl. Phys. Lett. {\bf 75}, 3171 (1999).
\bibitem{temp}The pinning force of AJ vortices is the maximum gradient of the 
magnetic energy $f(x)=-\phi_0\partial_x H(x)/4\pi$, where  
$H(x)\approx B+\Delta H\cos(2\pi x/a)$ is the local field of the A vortex lattice along GB, 
$\Delta H=(\phi_0/\pi\lambda^2)\exp(-2\pi u/a)$ at $H\gg H_{c1}$ and $u\approx a$. 
Thus, $f_m=\phi_0\Delta H/2a$, $\delta\beta(x) = c\partial_x H(x)/4\pi J_b=\beta_1\cos(2\pi x/a)$, 
and $\beta_1=c\Delta H/2aJ_b=4(H/H_0)^{1/2}\exp(-2\pi u/a)$. 
This yields the pinning correction to the resistivity, $R(J)-R_F\sim -R_FJ_{gb}^2/J^2$ for 
$J\gg J_{gb}$. Here $\beta_1\ll 1$ in a wide field range $H<2\times 10^4H_0$, where GB is 
transparent to local currents $J_v=c\partial_x H(x)/4\pi < J_b$ of A vortices. 
\bibitem{kes}R. Besseling, R. Niggerbrugge, and P.H. Kes, \prl {\bf 82} 3144 (1999).
\bibitem{mod}A. Gurevich and E.A. Pashitskii, \prb {\bf 57}, 13878 (1998). 
\bibitem{cai}X.Y. Cai {\it et al.}, \prb{\bf 57}, 10951 (1998).  
\bibitem{kunchur}M.N. Kunchur {\it et al.}, \prl{\bf 84}, 5204 (2000).

\end{references}
\end{document}